\documentclass[draft,jgrga]{agutex}

\usepackage[dvips]{graphicx}
\setkeys{Gin}{draft=false}

\usepackage{lineno}
\linenumbers*[1] 

\begin{document}
\authorrunninghead{Dorville et al.}

\titlerunninghead{Magnetopause: BV technique}

\authoraddr{Corresponding author: N. Dorville, LPP, Ecole Polytechnique, CNRS, UPMC, Universit\'e Paris Sud, Palaiseau, France.(nicolas.dorville@lpp.polytechnique.fr)}

\title{BV technique for investigating 1-D interfaces}

\authors {Nicolas Dorville, {\altaffilmark{1}} 
 G\'erard Belmont, {\altaffilmark{1}}
 Laurence Rezeau,  {\altaffilmark{1}}
 Nicolas Aunai,  {\altaffilmark{2}}  
 Alessandro Retin\`o, {\altaffilmark{1}} 

 }

\altaffiltext{1}{LPP, Ecole Polytechnique, CNRS, UPMC, Universit\'e Paris Sud, Palaiseau, France}
\altaffiltext{2}{NASA Goddard Space Flight Center, NPP - Space Weather Laboratory (674), Greenbelt, MD, USA}

\begin{abstract}

To investigate the internal structure of the magnetopause with spacecraft data, it is crucial to be able to determine its normal direction and to convert the measured time series into spatial profiles. We propose here a new single-spacecraft method, called the BV method, to reach these two objectives. Its name indicates that the method uses a combination of the magnetic field (B) and velocity (V) data. The method is tested on simulation and Cluster data, and a short overview of the possible products is given. We discuss its assumptions and show that it can bring a valuable improvement with respect to previous methods. 

\end{abstract}

\begin{article}

\section{Introduction}

 The Earth magnetopause is the outer boundary of the terrestrial magnetosphere. Outside of this boundary, the magnetosheath plasma is the shocked solar wind plasma, $i.e.$ cold and dense, with a magnetic field direction essentially determined by the solar wind one. Inside of it, the magnetospheric plasma is comparatively hot and tenuous, with a magnetic field direction essentially determined by the planetary one. Experimentally, investigating the magnetopause structure by spacecraft measurements is made difficult by the fact that the boundary is not steady: it can be shaken by the variations of the solar wind pressure, and perturbed by different kinds of waves, incident body waves as well as surface waves. It can also be locally and temporarily the place of different surface instabilities, implying or not magnetic reconnection, such as Kelvin-Helmholtz, Rayleigh Taylor or tearing instabilities (\textit{Hasegawa et al}, 2012). 

Two informations are crucial to investigate the magnetopause nature: 1) accurately determine the direction of its normal with respect to the magnetic field (in a strictly stationary configuration, having the normal magnetic field $B_n$ null or not have quite different consequences on the physical nature of the layer, even if the non null $B_n$ is small) and 2) determine an approximate spatial coordinate along the normal, to be able to draw the spatial profiles of the different relevant parameters, in the boundary frame, $i.e$ independently of the velocity at which these profiles are traversed by the spacecraft. 

Several methods have been developed for both of these two purposes.

To study the large scale shape of the boundary, its motion and its orientation, multi-spacecraft methods have been developed, particularly for the ESA Cluster mission (\textit{Paschmann and Daly}, 1998 and 2008). These methods are essentially based on timing differences between spacecraft and all rely on strong assumptions on the boundary: its form (plane or slightly curved at the scale of the spacecraft tetrahedron), its stationarity (constant profile and width, hereafter CTA for ``Constant Thickness Approach'', (\textit{Haaland et al}, 2004), or its velocity with respect to the spacecraft (hereafter CVA for ``Constant Velocity Approach'', \textit{Russell et al} 1983). Others are single spacecraft: they also rely on assumptions on the boundary properties such as planarity and stationarity, but they use in addition theoretical knowledge on the measured physical quantities, such as conservation laws. When using in particular the magnetic field data, the MVAB method (Minimum Variance Analysis on the magnetic field $\textbf B$, \textit{Sonnerup and Scheible}, 1998) takes advantage that div$(\textbf B)=0$, which draws $B_n = cst$ in the 1-D case. Its variant MVABC (C for corrected) adds the constraint   $B_n = 0$, using the additional information that the magnetopause normal component $B_n$ is generally close to zero, if not strictly zero.  This allows to handle cases when two components are nearly constant and not a single one ($i.e.$ when two eigenvalues of the variance-covariance matrix are small).
When the magnetopause can be supposed 1-D and stationary but when its thickness is small, making the kinetic effects non negligible with respect to the MHD ones, the experimental profiles have to be compared with the kinetic models of the tangential layers that can be found in the literature (see \textit{de Keyser and Roth}, 1998, for a review of the first models of this kind, and \textit{Belmont et al.}, 2012, for the most recent one). The experimental method developed in this paper should enable to perform such comparisons.
When the magnetopause layer cannot be supposed 1-D, other methods are needed. Some have been developed to reconstruct the magnetopause structure, supposing it is 2-D and stationary, and that it respects MHD equations: these are the Grad-Shafranov reconstruction methods (see \textit{Hasegawa et al}, 2004, for long-duration reconstruction). A review and discussion of short- and long- duration methods is made in \textit{De Keyser} (2006). Experimentally, it is often difficult to decide whether the 1-D or the stationary hypothesis has to be questioned first. Future comparison between the results of the reconstruction methods and those of the method proposed in this paper should be interesting in this respect. 

To find an approximate normal coordinate allowing to investigate the internal structure of the layer and to determine profiles across it, other methods have been developed independently, introducing the notion of "transition parameter" (\textit{Lockwood and Hapgood}, 1997). These methods can be used with single-spacecraft data. They also rely on assumed magnetopause properties, and they have been based hitherto on the variations in density and temperature of the electron population. This of course limits the temporal resolution of the method -and consequently its spatial one- to the electron experiment resolution. 

We propose here a new single spacecraft method, referred hereafter as ``BV'' to show that the magnetic field and the flow velocity data are used simultaneously, to analyze magnetopause-like interfaces. It combines the two previous types in such a way that it allows to determine in the same operation the magnetopause normal with an improved accuracy  and a transition parameter with an improved time resolution and expectingly closer to a real spatial coordinate. Fitting the magnetic field hodogram with a prescribed form, which is here an elliptical arc, allows to determine the normal direction with a fairly good accuracy. In addition, the angle $\alpha$ characterizing the position on the elliptical arc provides a reliable transition parameter inside the current layer, which can be viewed as a proxy for a normalized coordinate in the normal direction. On the other hand, as soon as the normal direction is known, the velocity measurements give a non-normalized normal coordinate, which is just the integral of the normal flow velocity $u_n$. It can give, in particular, a fairly good estimate of the physical width of the layer whenever the measured velocity should be in most cases dominated by the motion of the boundary. Using simultaneously the magnetic and velocity measurements just consists in imposing that the normal coordinate determined by the only velocity measurements is proportional to the transition parameter coming from the only magnetic measurements. Since the integral of $u_n$ is very sensitive to the normal direction, this enables to improve the determination of this direction with respect to the only magnetic one, while the time resolution of $y(t)$ remains approximately the magnetic one, which is much better than the velocity resolution.  

Section 2 presents the principles of the BV technique, and section 3 the different validation tests performed. The method allows to draw spatial profiles of any physical parameter across the magnetopause boundary. Examples of such profiles are presented in section 4, before discussing the interest and the limitations of the BV method and concluding in section 5.

\section{Principles of the method}

\begin{figure}[h]

\noindent\includegraphics[width=20pc]{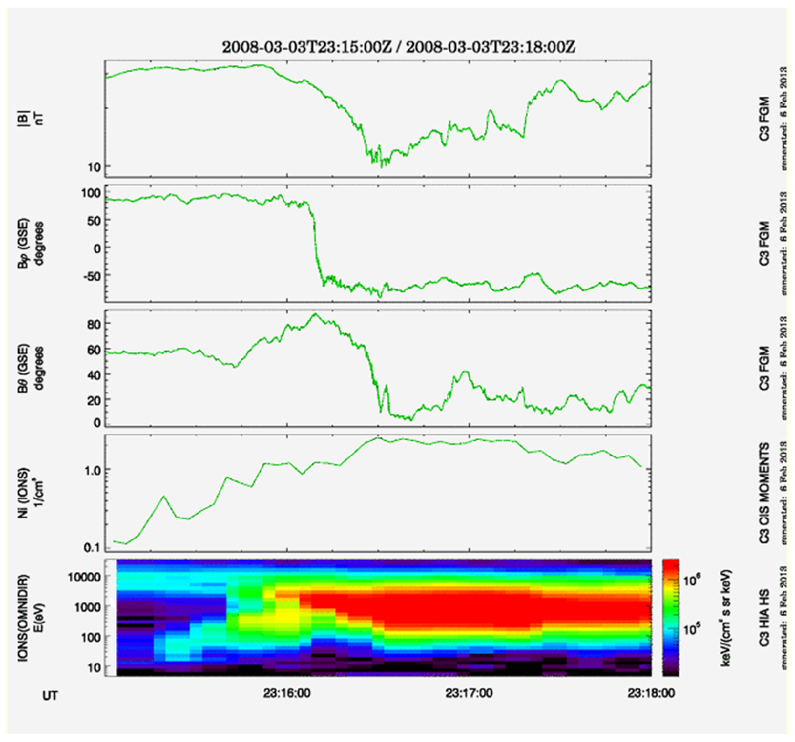}
\caption{Density, energy spectrogram and magnetic field observed by Cluster C1 around 23h16 on 03/03/2008. The jump of density, change in plasma energy composition and rotation of magnetic field show that the satellite is crossing the magnetopause.}
\label{CAA profiles at the crossing}

 \end{figure}

As the previous equivalent methods, the basic assumption of the BV technique is that, apart from oscillating perturbations, the boundary is sufficiently one dimensional and stationary at the scale of the spacecraft crossing. To explain the principles of this method, we use here a set of Cluster data on March, 3rd, 2008, when Cluster C3 encounters the magnetopause around 23:16, as it can be seen on Fig.~\ref{CAA profiles at the crossing} from the transition in the energy composition of the plasma, the density gradient and the rotation of magnetic field observed. The method uses principally the magnetic field data (\textit{Balogh et al.}, 1997). In subsection 2.1 we describe how we obtain an initial guess with only magnetic field data. Subsection 2.2 then explains the BV method itself, which combines magnetic field and ion velocity data.

 \subsection {Initialization with the only magnetic field data}
\begin{figure}[h]
\noindent\includegraphics[width=20pc]{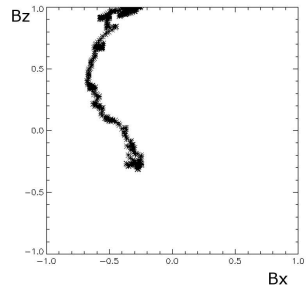}

 \caption{Hodogram of the magnetic field in the tangential MVABC plane for the Cluster C3 magnetopause crossing of 03/03/2008.}

 \label{hodogram LMN}
 \end{figure}

\begin{figure}[h]

\noindent\includegraphics[width=20pc]{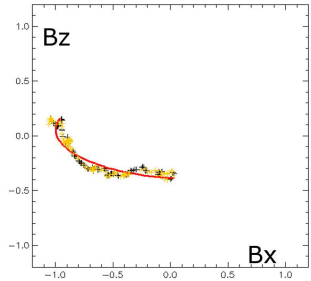}

 \caption{Initialization fit of the hodogram in initialization frame for the Cluster C3 magnetopause crossing of 03/03/2008. The data is in black, the selected points in yellow and the initialization fit in red.}

 \label{first fit}

 \end{figure}

In order to correctly initialize the minimization process of the complete BV method, involving magnetic field and ion velocity data, it is necessary to perform first an initialization stage, which provides an approximated frame and a first elliptical fit. This stage uses only the magnetic field data. It is done itself in several steps. The first step consists in finding a first approximation of the normal direction via a MVABC technique (\textit{Sonnerup and Scheible}, 1998). Fig.~(\ref{hodogram LMN}) shows the tangential hodogram derived by this method. 
In this example as in many other observations (\textit{Panov et al}, 2011), we observe a C-shaped hodogram, which can be fitted by an elliptical model. Although the general concept of the BV method is valid for any 1D layer, its present implementation is conceived for such kind of hodograms. Further generalization to more complicated hodograms (in particular for the S-shaped hodograms described in \textit{Panov et al}, 2011) is of course always possible. The second step consists in selecting the ``magnetic ramp'', $i.e.$ the interval of data where the  gradient of $ B_L $ is located. We then further select the data points by choosing only a sample of "representative points" among them. This step has a double purpose: eliminate the perturbations that can be considered as ``noise'', and make the different parts of the crossing equally represented in the statistics, even if the spacecraft does not spend the same time in these different parts. First we roughly eliminate the perturbations by discarding all points too far from the mean trajectory of the hodogram, and we represent each too close packet of points by only one single point. An elliptic fit and a new reference frame are derived from these points, using a Powell algorithm. The points selected in this way and the correspondent fit in the new frame are shown in Fig.~(\ref{first fit}). Then, the second goal is achieved by keeping a constant number of points in each $\alpha$ slice, which corresponds to the hypothesis (to be justified in next section) that $\alpha$ varies linearly with $y$. A new elliptic fit and approximated frame are then obtained, which provides a fine initialization for the BV method itself.

 \subsection {Simultaneous use of magnetic and velocity data}

The above stage has given an initial guess for the BV method regarding 1) the normal direction, and  2) the parameters describing the elliptic hodogram. The main part of the method then consists in using the temporal information $\textbf B(t)$, together with the velocity measurements from the Hot ion analyser experiment (\textit{R\`eme et al}, 1997). Going back to the totality of the $\textbf B$ data points, one minimizes the distance between them and the elliptical model $\textbf B(y)$, the function $y(t)$ being the integral of the normal velocity $u_n$. We therefore assume that this velocity is dominated by the layer velocity, $i.e.$ that the normal velocity in the layer frame is zero or negligible. The minimization is done with respect to the three angles that characterize the rotation of the ellipse proper frame and to the parameters of the elliptic hodogram, initialized previously, using the same Powell algorithm as above. The distance to be minimized  is:
\begin {equation}
\sum{\sqrt{(B_{dx}-B_{mx})^2+(B_{dy}-B_{my})^2+(B_{dz}-B_{mz})^2}}
\end {equation}
Where $\textbf B_{d}$ represents the data points and $\textbf B_{m}$ represents the model. This model is given by:
\begin{eqnarray}
B_{mx} &=& B_{x0}\cos{\alpha}
\\
B_{my} &=& B_{y0}
\\
B_{mz} &=& B_{z0}\sin{\alpha}
\end{eqnarray}
with:
\begin {equation}
\alpha = \alpha_{1}+(\alpha_{2}-\alpha_{1}) \  y/y_{max}, 
\end {equation}
 $y$ being the position deduced from the normal velocity integral. The magnetic field data and velocity data are obtained from prepared data by a rotation of M($\theta,\phi,\chi$). The parameters of the fit are $\theta,\phi,\chi,B_{x0},B_{y0},B_{z0},\alpha_{1},$ and $\alpha_{2}$.

\begin{figure}[h]

\noindent\includegraphics[width=20pc]{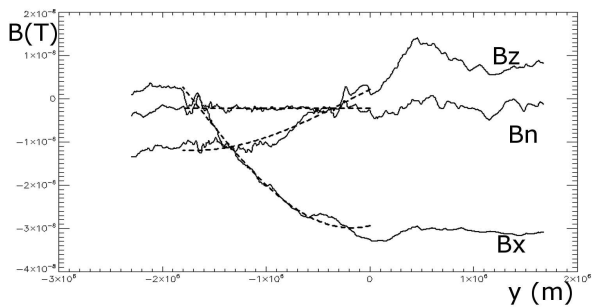}

 \caption{Fit (dashed) of the three components of the magnetic field (solid lines) along the normal coordinate for the Cluster C3 magnetopause crossing of 03/03/2008.}

 \label{fitcomponents}
 \end{figure}

This final stage provides all the needed outputs: the normal direction, the spatial position $y(t)$ along this normal (measured directly in physical units, providing in particular the layer thickness in km), and the fit of magnetic field, as illustrated, for example, on Fig.~(\ref{fitcomponents}). Here the computed magnetopause thickness is 1800~km and the linear Pearson correlation coefficients of the fit of $ B_x $ and $ B_z $ are 0.99 and 0.95. The spatial position $y$ is then extrapolated linearly outside the boundary, in order to plot approximated profiles of any plasma parameter on scales larger than the ramp region if necessary.

\section{Validations of the method}

Having presented how the BV method works in the previous section, we will now explain what led us to this way of proceeding and what are the different validation tests we have performed. We will discuss first the validity and the limitations of the hypotheses done, and discuss afterward the consistency of the obtained results.  We used three different tools to develop and validate the method: - a simple code to generate artificial magnetic field data, - a hybrid simulation of an asymmetric reconnection layer (\textit{Aunai et al}, 2013b), - and real data from the Cluster mission, especially a 2008 low latitude crossings list compiled by N. Cornilleau-Wehrlin. 

\subsection{Hypotheses: elliptical shape and linear angular velocity}

The first new assumption of the method, with respect to previous single spacecraft data analysis methods, is the elliptic shape of the tangential magnetic field hodogram, the simplest model geometry to describe C-shaped hodograms. This elliptical shape is indeed consistent with a simple generalization of the circular model $\textbf B (y)$ proposed by (\textit{Panov et al}, 2011):
\begin{eqnarray}
\frac{B_L}{B_{L0}} &=& \tanh(y/L)
\\
\frac{B_M}{B_{M0}}  &=& \frac{1}{\cosh(y/L)}
\end{eqnarray}
These formulas imply in particular that $B_L^2 / B_{L0}^2 + B_M^2 / B_{M0}^2 = 1$, which can be a test of the elliptical shape.

\begin{figure}[h]

\noindent\includegraphics[width=20pc]{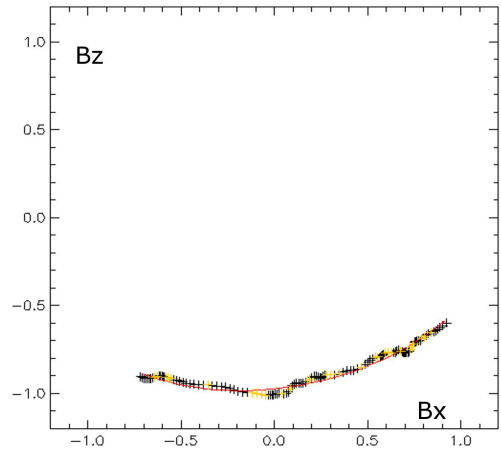}

 \caption{Fit of the hodogram of the magnetic field in the simulation for a virtual satellite crossing far from X point in the numerical simulation. The data is in black, the selected points in yellow and the fit in red.}

 \label{fitaunai}
 \end{figure}

The efficiency of the method can be tested first on a numerical simulation of reconnection (\textit{Aunai et al}, 2013b), far from the X point. Its applicability is not obvious in this case, since, before the development of the reconnection pattern, the initial condition is purely tangential, without any rotation. Nevertheless, the Hall effect creates a self-consistent out-of-plane magnetic component during the reconnection process, which, in the considered asymmetric configuration (asymmetric in density and temperature and coplanar and antisymmetric in magnetic field), results in a C-shaped hodogram if looked between the separatrices.  Fig.~(\ref{fitaunai}) shows the magnetic field in the interval that corresponds to the gradient of $B_L$. The error is here less than 2 percent. We have checked that this good accuracy is kept as long as the crossing considered is not too close to the X-point, which is generally the case for crossings of reconnected magnetopause or to the limits of the simulation.

\begin{figure}[h]

\noindent\includegraphics[width=20pc]{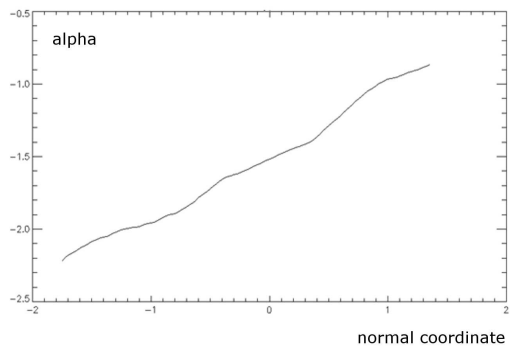}

 \caption{Angular position on the ellipse along the normal direction in the simulation for a virtual satellite crossing far from X point.}

 \label{alpha(y)aunai}
 \end{figure}

Concerning the analytical form of $\alpha(y)$, we also checked the validity of the linear hypothesis in the same simulation study. Fig.~(\ref{alpha(y)aunai}) shows how $\alpha$ varies as a function of the normal coordinate $y_s$ of the simulation . We observe that, apart from weak periodic variations, the linear form is well satisfied. It is worth explaining that the weak periodic departures from the linear variations (which can be well described by the three of four first terms of a Fourier transform) can indeed be accounted for in the minimization procedure, but it would increase the number of free parameters and drastically affect the convergence of the minimization process.

\subsection{Consistency of the results and limitations}
 
Regarding the consistency of the results, the first test consists in running the first part of the method (identification of the ellipse and of its proper frame) on a magnetic field that is artificially generated with an elliptic hodogram. Such artificial data have thus been constructed with the same analytical formulas as those of the program, then turned on a random frame, and added with a random Gaussian noise centered on the  signal, with a relative amplitude up to 50\%. The result is that the method always allows to find the good initial normal direction with at least 5 significant numbers, as well as the right ellipse parameters, whenever the noise does not exceed 30\%.

The second test consists in using the above numerical simulation (\textit{Aunai et al}, 2013b) to mimic a real magnetopause crossing. In order to make the method work, we must modify the simulation results in a way that makes it likely closer to most real magnetopause crossing: we multiply the tangential velocities by a large factor ($\approx 10$). Thanks to this change, the tangential velocities get a much larger contrast than the normal ones, which is necessary for the program convergence. It must be noted that such a  contrast of the tangential velocities does generally exist at the magnetopause, since the tangential velocity change is generally of the order of  a few 100 km/s, while the normal one (in the spacecraft frame) is generally about ten times smaller and varies very little. In order to focus on the reconnection process freely of any KH instability, the simulation did not include such a velocity shear.  Furthermore, the normal velocity of the virtual spacecraft considered with respect to the boundary, has to be chosen large enough with respect to the normal velocities in the boundary frame. This is also, as already mentioned, a reasonable hypothesis for a real magnetopause crossing.

Under these assumptions, we get normals with an angular precision oscillating between 0 and 5 degrees (with the corresponding errors on the shape of the tangential hodogram) and 0-5\% errors on the $y$ parameter (and derivative), which corresponds to the internal velocity and the approximations on $\alpha (y)$.

The result is not changing as long as the virtual spacecraft crosses the simulation far enough (several $d_i$) from the X point, where the 2D effects are not dominant. In these cases, the precision of the MVABC method is of the same order, (slightly better or worse, depending on the cases), because $B_n$ is actually very close to 0.

\begin{figure}[h]

\noindent\includegraphics[width=20pc]{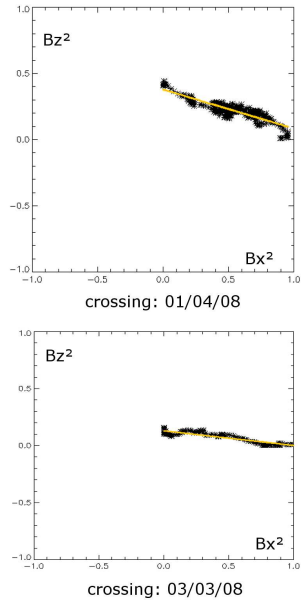}

 \caption{Hodogram of the squared components of magnetic field in the LM plane for two Cluster C3 magnetopause crossings of April, 1st 2008 and March, 3rd, 2008.}

 \label{champscarresLMN}
 \end{figure}

Regarding real Cluster data, the measurements show more perturbations, but the variations of the field value around the mean ellipse are still around 5 percent for most C-shaped hodograms. A good test for the elliptical shape is to plot  $ B_z^2(B_x^2) $, that should be linear for a tangential ellipse. Fig.~(\ref{champscarresLMN}) shows this plot for two magnetopause crossings on 03/03/2008 and 04/01/2008. It shows that the elliptical shape is a good approximation.

It is clear, from the the tests on the numerical simulation, that the BV method has limitations related to the necessary contrast between the normal and tangential component profiles. When applied to real Cluster data, these limitations may have, in some occasions, consequences on the results obtained. We will discuss these limitations in the conclusion section. It is to be noted however that these limitations are based on assumptions which are different -and generally weaker- than those of the other single-spacecraft methods such as MVAB or MVABC. 

\begin{figure}[h]

\noindent\includegraphics[width=20pc]{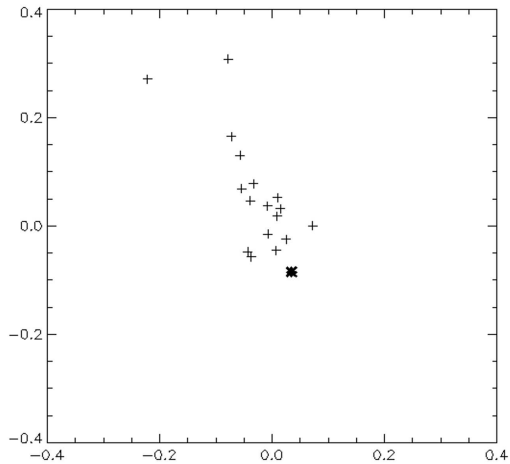}

 \caption{Several single and multi-spacecraft methods normal direction positions in the plane perpendicular to the MVABC mean normal for a benchmark case from (\textit{Haaland et al}, 2004). The star represents the result of BV method. The distance from the origin in this plane corresponds to the  $\sin\theta$, where $\theta$ is the angle from the mean normal direction. }

 \label{eye}
 \end{figure}

We will present a detailed study on a case (\textit{Dorville et al, 2013}), where the BV method leads to a better understanding and more precise results than MVAB(C).  When all the methods are confidently applicable, the results seem to be consistent with each other and with the theoretical knowledge. We show on Fig.~(\ref{eye})"oeil" a reproduction of a figure from (\textit{Haaland et al}, 2004) corresponding to a benchmark case where different methods have been used. The center of the figure is the mean MVABC normal and other single and multi-spacecraft methods are represented in a polar plot in the plane perpendicular to this normal. The result of the BV method on C1 spacecraft is indicated by a star. The figure shows that, if the result is different from other methods, it is inside the dispersion range of the points. The thickness of the layer always stands between a few hundreds of kilometers and a few thousands, which is consistent with literature, the tangential velocities being generally one order of magnitude larger than the normal one (in the spacecraft frame). The normal magnetic fields always stand between 0 and 20 nT, the non null values being reliable and quantitative indications of a connected boundary, which could hardly be obtained previously.   

\section{Products of the method}

As explained above, the first main direct product of the method is an accurate determination of the direction normal to the boundary, leading to reliable values of the small components $B_n$ and $u_n$ of the magnetic field and the flow velocity across the boundary. The second direct product is the determination of a spatial coordinate $y(t)$ allowing to draw any plasma parameter profile against the spatial position $y$ from their temporal measurement. The magnetopause layer thickness is also an interesting by-product deriving directly from the two preceding ones. 

\begin{figure}[h]

\noindent\includegraphics[width=20pc]{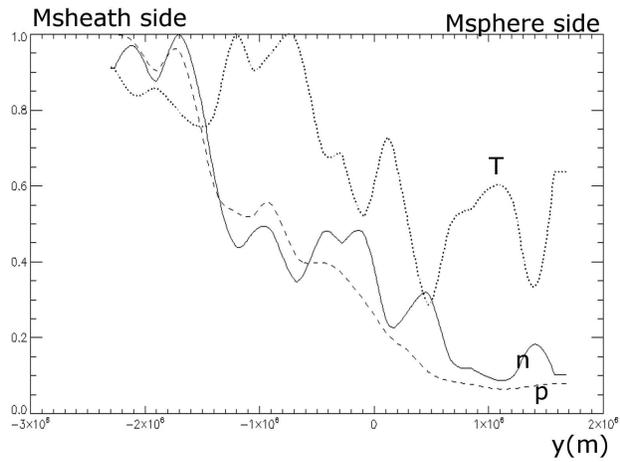}

 \caption{Evolution with normal position of density, pression and temperature measured by Cluster C3 for the 03/03/2008 crossing.}

 \label{CIS}
 \end{figure}

\begin{figure}[h]

\noindent\includegraphics[width=20pc]{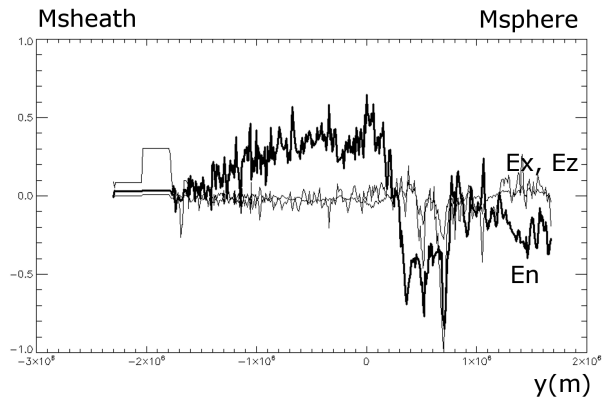}

 \caption{Evolution with normal position of the Electric field measured by Cluster C3 for the 03/03/2008 crossing.}

 \label{Efield}
 \end{figure}

\begin{figure}[h]

\noindent\includegraphics[width=20pc]{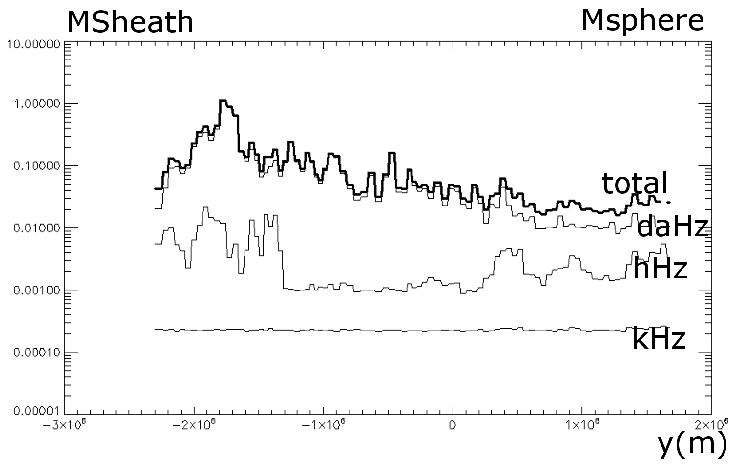}

 \caption{Evolution with normal position of the spectral power density for magnetic field measured by Cluster C3. The total spectral power density is the thick line, as the daHz hHz and kHz frequency ranges are also represented on the same scale.}

 \label{staffdata}
 \end{figure}

Examples of $y$ profiles are presented in Fig.~(\ref{CIS}) for the crossing of 03/03/08. Here we see the characteristic jump of density at the magnetopause, but no temperature jump, the pressure evolving like the density. For the different crossings that we investigated,  we could often observe clear differences concerning the locations of the particle gradients with respect to the magnetic field rotation. In a companion paper, we will present an interesting case study where the BV method can bring new information about the nature of the magnetopause.
\\
In Fig.~(\ref{Efield}) the normal electric field obtained with the EFW experiment (\textit{Gustafsson et al},(2001)) and the tangential components are shown for the same 03/03/2008  crossing. We see that the maximum variance is on the normal electric field, as expected by theory, and quite constant tangential electric fields, which confirms that the normal direction found is a good one. 
Fig.~(\ref{staffdata}) shows the profiles of magnetic field spectral power density obtained with the STAFF experiment (\textit{Cornilleau-Wehrlin et al}, 2003)for different frequency ranges. One can observe that the source of waves lies in the magnetosheath and that the depth of penetration depends on the frequency, the lowest frequencies penetrating deeper toward the magnetospheric side. 

This ability to get spatial profiles of all the quantities in the boundary is a key to a better understanding of the physical nature of the magnetopause.

\section{Discussion and conclusion}

We have presented the new BV method to analyze the structure of the magnetopause boundary layer, using spacecraft data. It combines the magnetic field and velocity measurements of one single spacecraft and permits to find the normal direction and a good resolution on a spatial coordinate to resolve small scale variations inside the layer. Using it, we are able to study the internal structure of the layer, for any of the physical quantities measured on board. The method works on simulation and generated data, and its assumptions can be verified on Cluster crossings.

It is worth observing the conditions of validity of the BV method are not the same as the other single spacecraft methods such as MVAB, and that they are in general less restrictive. 
In MVAB, one needs to discriminate $B_N$ and $B_M$, which fails systematically in structures as shocks, and often at the magnetopause since this one is often quasi-coplanar. MVABC has the same condition of validity, with the additional problem that it cannot be used for determining $B_n$ since this component is supposed null. 
In the BV method, one needs to discriminate the two couples of data sets: ($B_N,\  V_N$) and ($B_M,\  V_M$). This is clearly a weaker condition since, even if $B_N$ and $B_M$ are nearly constant, the differences between $V_N$ and $V_M$, (profiles and/or orders of magnitude) are generally sufficient to guarantee a correct operation. The difficulties can only arise when  not only $B_N$ and $B_M$ are indistinguishable (mean jump much smaller than noise), but also $V_N$ and $V_M$. 

Contrary to the multi-spacecraft timing methods, the BV method can also handle cases when the boundary is shaken with a non trivial normal velocity evolution (which seems frequent). When this evolution is non negligible between two spacecraft crossings, the timing methods obviously fail.

The BV method however brings a new limitation: although one works essentially with magnetic field data, a sufficiently long crossing is needed (at least three or four velocity measurement points inside the crossing)to make efficient the contribution of the velocity data. We are therefore not able to analyze as many crossings as the other methods.

With the proposed method, the structure of the magnetopause should be now open to more detailed investigations. Some examples of spatial profiles have been given in section 4. The method is used in a companion article, for an atypical magnetopause case study giving new insight on this structure. 

\end{article}

\begin{acknowledgments}
The authors would like to thank N Cornilleau-Wehrlin for fruitful discussion and her help to work with Cluster data and detect magnetopause crossings, and the CAA and all Cluster instruments teams for their work on Cluster data.
\end{acknowledgments}


%

\begin{thebibliography}{}
\bibitem[{\textit{Aunai et al}(2011)}]{Aunai}
Aunai, N., G. Belmont, and R. Smets (2011), Proton acceleration in antiparallel collisionless magnetic reconnection: Kinetic mechanisms behind the fluid dynamics, J. Geophys. Res., 116, A09232, doi:10.1029/2011JA016688.
\bibitem[{\textit{Aunai et al}(2013b)}]{Aunai2}
Aunai, N., et al. Comparison between hybrid and fully kinetic models of asymmetric magnetic reconnection: Coplanar and guide field configurations. Physics of Plasmas 2013; 20: 022902.   
\bibitem[{\textit{Balogh et al}(1997)}]{Balogh}
Balogh, A., Dunlop, M. W., Cowley, S. W. H., Southwood, D. J., Thomlinson, J. G., and the Cluster magnetometer team: The Cluster magnetic field investigation, Space Sci. Rev., 79, 65, 1997
\bibitem[{\textit{Cornilleau-Wehrlin et al}(2003)}]{Cornilleau-Wehrlin}
Cornilleau-Wehrlin, N., Chanteur, G., Perraut, S., Rezeau, L., Robert, P., Roux, A., de Villedary, C., Canu, P., Maksimovic, M., de Conchy, Y., Hubert, D., Lacombe, C., Lefeuvre, F., Parrot, M., Pin�on, J. L., D�cr�au, P. M. E., Harvey, C. C., Louarn, Ph., Santolik, O., Alleyne, H. St. C., Roth, M., Chust, T., Le Contel, O., and STAFF team: First results obtained by the Cluster STAFF experiment, Ann. Geophys., 21, 437-456, doi:10.5194/angeo-21-437-2003, 2003.
\bibitem[{\textit{De Keyser}(2006)}]{De Keyser}
De Keyser, J. (2006), The Earth magnetopause: reconstruction of motion and structure, Space Science Reviews 121: 225-235
\bibitem[{\textit{De Keyser et Roth}(1998)}]{De Keyser2}
De Keyser, J., and M. Roth (1998), Equilibrium conditions and magnetic field rotation at the tangential discontinuity magnetopause, J. Geophys. Res., 103(A4), 6653–6662, doi:10.1029/97JA03710.
\bibitem[{\textit{Dorville et al}(submitted 2013)}]{Dorville}
Dorville N., G.Belmont, L. Rezeau, R. Grappin, A. Retino, Rotational/ Compressional nature of the Magnetopause: Application of the BV technique on a magnetopause case study, submitted to JGR, 2013
\bibitem[{\textit{Gustafsson et al}(2001)}]{Gustafsson}
Gustafsson, G., Andr�, M., Carozzi, T., Eriksson, A. I., F�lthammar, C.-G., Grard, R., Holmgren, G., Holtet, J. A., Ivchenko, N., Karlsson, T., Khotyaintsev, Y., Klimov, S., Laakso, H., Lindqvist, P.-A., Lybekk, B., Marklund, G., Mozer, F., Mursula, K., Pedersen, A., Popielawska, B., Savin, S., Stasiewicz, K., Tanskanen, P., Vaivads, A., and Wahlund, J.-E.: First results of electric field and density observations by Cluster EFW based on initial months of operation, Ann. Geophys., 19, 1219-1240, doi:10.5194/angeo-19-1219-2001, 2001.
\bibitem[{\textit{Haaland et al}(2004)}]{Haaland}
Haaland, S., Sonnerup, B., Dunlop, M., Balogh, A., Georgescu, E., Hasegawa, H., Klecker, B., Paschmann, G.,Puhl-Quinn, P., Rème, H., Vaith, H., and Vaivads, A., 2004a, Four-spacecraft determination of magnetopause orientation, motion and thickness: comparison with results from single-spacecraft methods,Ann. Geophys.,22, 1347–1365
\bibitem[{\textit{Hasegawa}(2012)}]{Hasegawa}
Hasegawa, H. (2012), Structure and dynamics of the magnetopause and its boundary layers,
Monogr. Environ. Earth Planets,1, 71–119,doi:10.5047/meep.2012.00102.0071.
\bibitem[{\textit{Hasegawa et al}(2004)}]{Hasegawa2}
Hasegawa,H. et al (2004), Reconstruction of two-dimensional magnetopause structures from Cluster observations: verification of method, Annales Geophysicae (2004) 22: 1251–1266.
\bibitem[{\textit{Lockwood et Hapgood}(1997)}]{Lockwood}
Lockwood, M., and M. A. Hapgood, How the magnetopause transition parameter works, Geophys. Res. Lett., 24, 373-376, 1997.
\bibitem[{\textit{Panov et al}(2011)}]{Panov}
Panov, E. V., A. V. Artemyev, R. Nakamura, and W. Baumjohann (2011), Two types of tangential magnetopause current sheets: Cluster observations and theory, J. Geophys. Res., 116, A12204, doi:10.1029/2011JA016860.
\bibitem[{\textit{Paschmann et al}(1998)}]{Paschmann}
Paschmann, G., and Daly, P., W., 1998, Analysis Methods for Multi-spacecraft data, no. SR-001 in ISSI Scientific Reports, ESA Publ. Div., Noordwijk, Netherlands.
\bibitem[{\textit{Paschmann et al}(2008)}]{Paschmann2}
Paschmann, G., and Sonnerup, B. U. Ö., 2008, Proper Frame Determination and Walén Test, in Multi-spacecraft Analysis Methods Revisited, edited by G. Paschmann and P. W. Daly, no. SR-008 in ISSI Scientific Reports, ESA Publ. Div., Noordwijk, Netherlands.
\bibitem[{\textit{R\`eme et al}(1997)}]{Reme}
R\`eme, H., Cottin, F., Cros, A., et al.: The Cluster Ion Spectrometry (CIS) Experiment, Space Sciences Review 79, 303-350, 1997
\bibitem[{\textit{Russell et al}(1983)}]{Russell}
Russell, C. T., Mellot, M. M., Smith, E. J., and King, J. H., 1983, Multiple spacecraft observations of interplanetary shocks: Four spacecraft determination of shock normals,J. Geophys. Res.,88, 4739–4748.
\bibitem[{\textit{Sonnerup et al}(1998)}]{Sonnerup}
Sonnerup, B. U. Ö., and Scheible, M., 1998, Minimum and Maximum Variance Analysis, in Analysis Methods for Multispacecraft Data, edited by G. Paschmann and P. W. Daly, no. SR-001 in ISSI Scientific Reports, chap. 8, pp. 187-196, ESA Publ. Div., Noordwijk, Netherlands.

 \end{thebibliography}
\end {document}